# Ne İzotoplarının Nükleer Yapı Özelliklerinin Nükleer Kabuk Modeli ile İncelenmesi

*[1]Serkan Akkoyun, [2]Tuncay Bayram

[1] Sivas Cumhuriyet Üniversitesi, Fen Fakültesi, Fizik Bölümü, Sivas, Türkiye, sakkoyun@cumhuriyet.edu.tr,
0000-0002-8996-3385

[2] Karadeniz Teknik Üniversitesi, Fen Fakültesi, Fizik Bölümü, Trabzon, Türkiye, t.bayram@ymail.com,
0000-0003-3704-0818

## Öz

Atom çekirdeklerinin nükleer yapılarını araştırma amacıyla kullanılan yaygın yöntemlerden birisi de, nükleer kabuk modelidir. Atom elektronlarının yörüngelere yerleşmesine benzer şekilde, nükleer kabuk modelinde de proton ve nötronların, Pauli dışarlama ilkesine uyarak çekirdek içerisinde yörüngelere yerleştiği düşünülmektedir. Bu yörüngeler, kendi aralarında gruplanarak kabukları meydana getirmektedir ki, bir kabuktaki tüm mümkün seviyelerin dolu olması durumunda, kabuğun kapalı olduğu söylenir. Kapalı kabuğa sahip atom çekirdekleri oldukça kararlıdırlar ve nükleer kabuk modeli hesaplamalarında bu çekirdeklerden fazla olan değerlik nükleonları hesaplamalara katılır. Bu çalışmada, $^{16}$O çekirdeği kapalı kabuk çekirdeği olarak ele alınarak, çift-çift Ne çekirdeklerinin nükleer yapılarını araştırmak için nükleer kabuk modeli kullanılmıştır. Tek parçacık yörüngeleri olarak $d_{5/2}$, $s_{1/2}$ ve $d_{3/2}$ ele alınarak, değerlik nükleonları arasındaki iki cisim etkileşmeleri için farklı parametre setleri kullanılmıştır. Sonuçlar birbirleriyle ve mevcut literatür değerleriyle karşılaştırılmıştır. Deneysel değerlere en yakın sonuçların, *usdb* ve *sdnn* parametre setleri ile elde edildiği görülmüştür.

*Anahtar kelimeler:* Nükleer kabuk modeli, nükleer yapı, Ne çekirdekleri

# Investigation of Nuclear Structures of Ne Isotopes by Nuclear Shell Model

*[1]Serkan Akkoyun, [2]Tuncay Bayram

[1] Sivas Cumhuriyet University, Faculty of Science, Department of Physics, Sivas, Turkey, sakkoyun@cumhuriyet.edu.tr
[2] Karadeniz Technical University, Faculty of Science, Department of Physics, Trabzon, Turkey, t.bayram@ymail.com

## Abstract

One of the common methods used to investigate the nuclear structures of atomic nuclei is the nuclear shell model. Similar to the placement of atomic electrons into orbits, in the nuclear shell model, protons and neutrons are thought to fill the orbits within the nucleus, following the principle of Pauli's exclusion. These orbits are grouped together to form shells, which are said to be closed if all possible places in a shell are full. Atomic nuclei with closed shells are very stable and valence nucleons that are more than these nuclei are included in the nuclear shell model calculations. In this study, the nuclear shell model was used to investigate the nuclear structure of even-even Ne nuclei by considering the $^{16}$O core as a closed shell nuclei. Single particle orbits $d_{5/2}$, $s_{1/2}$ and $d_{3/2}$ are taken into account and different parameter sets are used for two-body interactions between valance nucleons. The results were compared with each other and with current literature values. It was seen that the closest results to the experimental values were obtained with parameter sets of *usdb* and *sdnn*.

*Keywords:* Nuclear shell model, nuclear structure, Ne nuclei

*[1]Sorumlu Yazar: Sivas Cumhuriyet Üniversitesi, Fen Fakültesi, Fizik Bölümü, Sivas, Türkiye, sakkoyun@cumhuriyet.edu.tr

## 1. GİRİŞ

Atom çekirdeğinin etrafında var olduğu düşünülen atomun yörünge modelinde, elektronların bu yörüngelerde bulundukları düşünülmektedir. Elektronların bu yerleşimi, Pauli dışarlama ilkesine göre olur ve aynı kuantum sayısına sahip iki elektron aynı yörüngede asla bulunamaz. Her yörüngenin, kuantum sayıları ile ilişkili olarak alabileceği azami elektron sayısı vardır. Bu şekilde yörüngelerin elektronlarla dolması sonucunda, belirli bazı elektron sayısına sahip atomların, diğerlerine göre daha kararlı olduğu bilinmektedir. Bu atomlar, iyi bilindiği üzere soy gazlar olarak adlandırılmaktadır. Bu modele benzer bir modelin, atom çekirdeğinin içinde yer alan ve ortak adları nükleon olan proton ve nötronlara da uygulanabileceği görülmüştür. Nükleer kabuk modeli [1-5] olarak adlandırılan bu modelde nükleonlar, çekirdek içerisindeki yörüngelere Pauli ilkesine göre ayrı ayrı yerleşirler. Soy gazlara benzer olarak, bazı nükleon sayılı çekirdeklerin, diğerlerine göre daha kararlı oldukları gözlemlenmiştir ki bu sayılar nükleer fizikte sihirli sayılar (2, 8, 20, 28, 50, 82 ve 128) olarak adlandırılır [6, 7]. Hem nötron hem de protonu sihirli sayıda olan çift sihirli çekirdekler, küresel yapıda olup oldukça kararlıdır. Sihirli sayıya sahip tek parçacık yörüngelerinin, ardından gelen yörüngeler arasında fazla mesafe olması, yörüngelerin gruplanmış olarak bulunmalarına yol açar. Bu gruplara kabuk adı verilmekte olup, nükleer kabuk modeli ismi buradan gelmektedir. Şekil 1'de görülebilen bu kabuklar, içerdikleri yörüngelere göre isimlendirilirler. Son zamanlarda yapılan teorik ve deneysel çalışmalar, bazı kabuklar için mevcut sihirli sayılardan farklı yeni sihirli sayıların olabileceğine veya mevcutların sihirli olamayabileceğine işaret etmektedir [8].

Nükleer kabuk modeli hesaplamalarında, çift sihirli sayılı çekirdekler, öz (kor) çekirdek olarak ele alınarak, bundan fazla olan değerlik nükleonları hesaplamalara katılır. Öz çekirdekte *J=0* toplam açısal momentumu veren nükleonların hareket etmediği varsayılmaktadır. Buna göre bu nükleonların, özden çıkarak değerlik nükleonlarının içine dahil olmaları mümkün değildir. Değerlik nükleonlarının, özün hemen üzerindeki kabukta dağılmış olabileceği farz edilerek bu kabuk, model uzayı olarak ele alınır. Model uzayındaki nükleonlar, her bir yörüngede tüm kombinasyonlarda yerleşebilirler. Farklı yerleşimler, çekirdeğin farklı enerji seviyelerinin oluşmasına neden olur. Model uzayındaki yörüngelerin, yörünge kapasitelerinin artması ve değerlik nükleon sayısının artması, yapılacak olan hesaplamaları oldukça zorlaştırmaktadır. Bu çalışmada, $^{16}$O özü kullanılarak, *sd* kabuğunda yer alan çift-çift (hem proton hem nötron sayısı çift sayı) Ne çekirdeklerinin nükleer özellikleri araştırılmıştır. İncelenen Ne izotopları, öz dışında kalan 2 protona ve 0 ile 12 arasındaki çift sayıda nötrona sahiptir. Proton ve nötronlar, ayrı ayrı n olmak üzere, *sd* model uzayında her kombinasyonda dağılmıştır. Ne izotoplarının $2^+$, $4^+$ ve $6^+$ uyarılmış seviye enerjileri, $4^+$ enerjisinin $2^+$ enerjisine oranları, taban durumdan $2^+$ seviyesine olan indirgenmiş kuadrupol geçiş olasılıkları ve deformasyon parametreleri hesaplanmıştır. Farklı parametre setleri ile yapılan hesaplamalardan elde edilen sonuçlardan görüldüğü üzere, nükleonlar arasındaki etkileşimi tanımlayan *usdb* ve *sdnn* iki cisim matris elemanları setlerinin, deneysel değerlere daha yakın sonuçlar verdiği görülmüştür. Hesaplamalar için, Kshell kabuk modeli bilgisayar kodu kullanılmıştır [9].

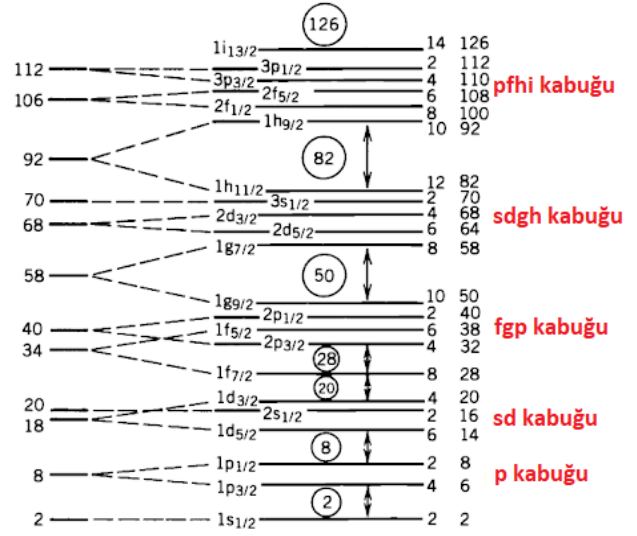

**Şekil 1.** Nükleer kabuk modeli yörüngeleri [10]

## 2. KABUK MODELİ HESAPLAMALARI

Nükleer kabuk modeli, atom çekirdeğinin düşük enerjili yapısını tanımlamak için en uygun araçtır [11]. Bu modelde nükleonların (proton ve nötronların), bağımsız bir merkezi potansiyel kuyusunda hareket ettiği varsayılmaktadır. Güçlü spin-yörünge etkileşiminin önemli bir bileşen olduğunun ortaya çıkmasının ardından, tek parçacık yörünge dizilimleri ve sihirli sayılar (2, 8, 20, 28, 50, 82 ve 126) bu etkileşmenin de dahil edilmesiyle Şekil 1'deki son şeklini almıştır [12, 13]. Hem proton hem de nötron bakımından bu sihirli sayıya sahip olan çekirdekler, diğerlerine göre daha kararlı yapıda ve küresel şekildedir. Çekirdeğin kabuk modeli, atomun kabuk modeline benzemekle birlikte, pek çok zorlukları da içermektedir. Bu zorluklardan ilki, atomun boyutlarının $10^{-10}$ m, çekirdeğinkinin ise $10^{-15}$ m mertebesinde olması nedeniyle, çalışma alanının darlaşmasıdır. Diğer bir zorluk ise, çekirdekte tek tip parçacık yerine, iki farklı parçacık türünün (protonlar ve nötronlar) bulunmasıdır. Atomun kabuk modelinde elektronlar, atomun yörüngelerinde merkezi bir potansiyel kuyusunda bağımsız olarak hareket eder. Merkezi potansiyel, çekirdeğin pozitif yükünden ve elektronların ortalama itici etkileşiminden kaynaklanmaktadır. Oysa çekirdeğin kabuk modelinde böyle bir merkezi potansiyel söz konusu değildir. Ayrıca bu modelde, elektronların kendi aralarında ya da çekirdek ile etkileşmelerinin Coulomb etkileşmesi ile tanımlandığı gibi açık bir tanımlama nükleonlar arasında mevcut değildir.

Nükleer enerji seviyelerinin hesaplanması oldukça zor bir iştir. Zorluğun ana nedeni, serbest protonlar ve nötronlar arasındaki etkileşimin, bir başka ifade ile güçlü nükleer etkileşmenin, doğasının yeterince iyi bilinmemesidir. Kapalı kabukların dışında birkaç değerlik nükleonun olduğu bir çekirdeği düşünürsek, seviyelerin enerjileri üç kısma ayrılabilir. Birincisi, kapalı kabukların bağlanma enerjisi, ikincisi değerlik nükleonlarının kinetik enerjileri ve öz çekirdeğin nükleonlarıyla etkileşimleri içeren tek nükleon enerjilerinin toplamıdır. Üçüncüsü ise, değerlik nükleonlarının birbirleriyle karşılıklı etkileşimidir. Bunların arasında kapalı kabukların bağlanma enerjilerini hesaplamak en zor olanıdır. Bunlar arasında hesaplanması en kolay olanı, değerlik nükleonları arasındaki etkileşimdir. Eğer bu değerlik nükleonları tek bir yörüngedeyse, sadece bu yörüngedeki nükleonlar arasındaki etkin etkileşimin matris elemanlarını bilmek yeterlidir. Değerlik nükleonları birkaç yörüngeye dağılmışsa, tek nükleon enerjileri arasındaki farklılıklara da ihtiyaç vardır ki bunlar genellikle deneysel verilerden alınabilir. Kabuk modeli hesaplamalarındaki en önemli noktalardan birisi de, değerlik nükleonları arasında kullanılacak etkin etkileşimin seçilmesidir [14].

Nükleonlar arasında bireysel etkileşmelerin bilinmemesinden kaynaklanan zorluktan ötürü, bu etkileşmeler yerine, diğer nükleonların oluşturduğu ortalama bir potansiyel (ortalama alan yaklaşımı) işin içine katılır. Böylelikle nükleer kabuk modeli kapsamında ele alınan problem, çekirdekteki tüm nükleonları hesaba katan çok-cisim problemi, sadece değerlik nükleonlarını hesaba katan birkaç-cisim problemine indirgenmiş olur.

$A$ tane nükleona sahip çekirdek için Hamiltonyen, Denklem 1'deki gibi yazılabilir.

$$H = \sum_{i=1}^{A} T_i + \frac{1}{2}\sum_{i,j=1}^{A} V_{ij} \qquad (1)$$

Burada $T_i$, her bir nükleonun kinetik enerjisi, $V_{ij}$ ise, nükleonlar arasındaki etkileşme potansiyelidir. Fakat nükleonlar arasındaki etkileşme açıkça tanımlı olmadığından dolayı, her bir nükleonun, diğerlerinin oluşturduğu ortalama bir potansiyelde hareket ettiğini varsayarak, Hamiltonyen Denklem 2'de verildiği gibi düzenlenebilir.

$$H = \sum_{i=1}^{A}[T_i + U_i] + (\frac{1}{2}\sum_{i,j=1}^{A} V_{ij} - \sum_{i=1}^{A} U_i) \qquad (2)$$

$$H = H_0 + H_{artık} \qquad (3)$$

Burada $H_0$, her bir nükleonun, ortalama bir potansiyel altındaki tek parçacık enerjisidir. Harmonik salınıcı, Wood-Saxon ya da Yukawa tipindeki gibi bir merkezi potansiyelin seçimi ile tek parçacık enerjileri belirlenebilir. Artık etkileşmeye ($H_{artık}$) ait iki-cisim matris elemanları ise, nükleonların karşılıklı etkileşmelerini temsil eder. Literatürde farklı yöntemlerle elde edilmiş matris eleman setleri mevcuttur. Bu matris eleman setleri, bir veya iki kabuğu kapsayacak şekildedir. İkincil kuantumlamada yaratma ve yok etme işlemcileri cinsinden Hamiltonyen, Denklem 4'deki halini alır.

$$H = \sum_{ij}^{A} \epsilon_i a_i^+ a_j + \frac{1}{4}\sum_{ijkl}^{A}(ij|V|kl)\, a_i^+ a_j^+ a_k a_l \qquad (4)$$

Burada $\epsilon_i$ değerleri, nükleonların öz çekirdekle etkileşmeleri olan tek parçacık enerjileridir. $j$ yörüngesinde yok edilen parçacık, $i$ yörüngesinde yaratılır. İki cisim etkileşme terimi ise, $k$ ve $l$ yörüngelerinde yok edilen parçacıkların $i$ ve $j$ yörüngelerinde yaratıldığını söyler. Nükleonlar için Slater determinantları ile tanımlanan çok-parçacık dalga fonksiyonu (Ψ) kullanarak Hamiltonyen denklemi (Denklem 5) çözülerek, çekirdeğin enerji seviyeleri hesaplanabilir.

$$H\Psi = E\Psi \qquad (5)$$

Çok parçacıklı sistemler için matris formalizminde gerçekleştirilen bu işlemlerde, model uzayının boyutu ve nükleon sayısı artıkça, Hamiltonyen matrisinin boyutları oldukça yüksek mertebelere ($10^{10}$) kadar çıkmaktadır. Özdeğerleri elde etmek için matrisler, Lanczos gibi uygun algoritmalar kullanılarak köşegenleştirilir ve çözüme ulaşılır. Bu amaçla, literatürde nükleer kabuk modeli hesaplamalarını yapmak için geliştirilmiş birçok bilgisayar kodu mevcuttur. Bunlara örnek olarak, Oxbash [15], Antoine [16], Nushell [17], Bigstick [18], Redstick [19] ve Kshell [9] verilebilir. Bu çalışmada gerçekleştirilen hesaplamalarda, Kshell kodu kullanılmıştır. Linux işletim sisteminde çalışan bu kod, Lanczos yöntemi kullanılarak M-şeması özdeğerlerinin de nükleer kabuk modeli hesaplamaları gerçekleştirmeyi sağlar. Çekirdeklerin enerji seviyeleri, spin ve izospinleri, manyetik ve kuadrupol momentleri, seviyeler arasındaki B(E2) ve B(M1) geçiş olasılıkları ve tek parçacık spektroskobik faktörleri kod ile $10^{10}$ boyutuna kadar hesaplanabilir. Çekirdeklerin deformasyonunu gösteren kuadrupol deformasyon parametresi ($\beta_2$) parametresi ise, Denklem 6 yardımı ile hesaplanabilir [20].

$$\beta_2 = \frac{4\pi}{3ZR_0^2}[B(E2)/e^2]^{1/2} \qquad (6)$$

Burada, $Z$, çekirdeğin proton sayısı, $R_0^2 = 0{,}0144A^{2/3}$ barn olup, $A$ ise çekirdeğin atom kütlesidir.

## 3. SONUÇLAR ve TARTIŞMA

Çalışmamızda, model uzayı olarak sd uzayını ele aldık. Bu uzay, $^{16}$O öz çekirdeğinin üzerinde yer alan, $d_{5/2}$, $s_{1/2}$ ve $d_{3/2}$ yörüngelerinden oluşmaktadır. Öz çekirdeğin 8 protonu ve 8 nötronu vardır. İncelenecek olan Ne çekirdeğinin proton sayısı 10 olduğundan, model uzayındaki değerlik proton sayısı 2'dir. Ayrıca, atom kütlesi 18 ile 30 arasındaki Ne çekirdekleri incelendiğinden, model uzayındaki değerlik nötronları 0 ile 12 arasındadır. Literatürde sd model uzayından kullanılmak üzere mevcut olan, *cw , cwh* [21], *hbumsd, hbusd* [22], *kuosd, kuosdh, pw* [23], *sdba, sdnn* [24], *w* [25], *usda* [26] ve *usdb* [27] iki cisim etkileşme matris eleman setlerini ayrı ayrı kullanarak, nötron sayısı 8 ile 12 arasındaki çift-çift Ne çekirdeklerinin nükleer özellikleri elde edilmiştir. Ne çekirdekleri içerisinde 20 ve 22 atom kütleli olanlar, kararlı olan çekirdeklerdir. Diğer çekirdeklerin yarı ömürleri ise en fazla saniyeler mertebesindedir. Öncelikle, bu çekirdeklerin ilk $2^+$ ve $4^+$ seviyelerinin enerjilerini kabuk modeli hesaplamaları ile elde ederek, bu enerjilerin birbirlerine oranları incelenmiştir. Şekil 2'de, Ne çekirdeklerinin ilk $2^+$ ve $4^+$ enerji seviyeleri, deneysel değerlere [28] en yakın sonuç veren 4 matris eleman seti için verilmiştir. Bu matris elemanları, *usd* tipi matris elemanları olup, *w* matris eleman setinin yeni deneysel verilerle geliştirilmesi ile elde edilmiştir. *Sdnn* seti ise, en yenileri olan *usdb* setinin, yapay zeka ile bir miktar iyileştirilmesi ile türetilmiştir. Görüldüğü gibi, deneysel değerlere en yakın değerler, usdb ve *sdnn* kullanılarak elde edilmiştir. Buna göre, deneysel enerji değerlerinden sapmaların ortalama mutlak değerleri usdb, sdnn, usda ve w matris eleman setleri için sırasıyla, 161,56, 161,63, 192,67 ve 270,47 keV olarak elde edilmiştir. [18]Ne'den [28]Ne'ye kadar teorik hesaplamalardan elde edilen sonuçlar ile deneysel sonuçlar birbirine oldukça yakın iken, [28]Ne ve [30]Ne izotopları için teorik sonuçların deneyselden biraz uzaklaştığı görülmektedir. Bu ayrışma, özellikle $2^+$ seviyesinde daha fazladır. Ek olarak, literatürde mevcut olan deneysel verilerden birkaçı, tam kesinlikle belirlenememiştir. Bunlar, [26]Ne ile [28]Ne için $4^+$ seviyeleri ile [30]Ne için hem $2^+$ hem $4^+$ seviyeleridir. Bu çalışmada, tüm matris eleman setleri ile yapılan hesaplamaların yakın sonuçlar vermesi nedeniyle, [26]Ne ve [28]Ne için literatürde mevcut olan kesin olmayan sonuçlar desteklenmiştir. [30]Ne için $4^+$ seviyesi literatürdeki deneysel değerine yakın olarak hesaplanmasına karşın, bu çekirdeğin $2^+$ seviyesi için elde edilen tüm sonuçlar deneysel değerden oldukça uzaktır.

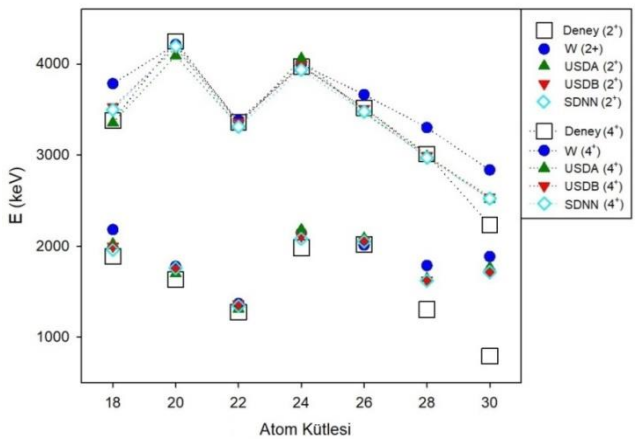

**Şekil 2.** Çift-çift Ne izotopları için *usd* tipi matris elemanları kullanılarak SM hesaplamalarından elde edilen ve deneysel $2^+$ ve $4^+$ enerji seviyeleri

Şekil 3 ve 4'de, Ne çekirdeklerinin ilk $2^+$ ve $4^+$ enerji seviyeleri, diğer matris eleman setleri için deneysel değerlerle birlikte gösterilmiştir. Şekilden de görülebileceği gibi, bu setler kullanılarak elde edilen enerji değerleri, usd tipi setlere göre deneysel değerlerden daha uzaktır. Buna göre, deneysel enerji değerlerinden sapmaların ortalama mutlak değerleri cw, cwh, hbumsd, hbusd, kuosd, kuosdh, pw ve sdba matris eleman setleri için sırasıyla, 292,67, 516,00, 618,40, 628,60, 376,00, 415,67, 316,07 ve 299,07 keV olarak elde edilmiştir. Bu sapmalar içinde usd tipi setlerin sapmalarına ve deneysel değerlere en yakın olanı, sdba seti olup en büyük sapma vereni ise, hbusd setidir. Yine [30]Ne izotopu için teorik sonuçların deneyselden oldukça uzak olduğu görülmektedir.

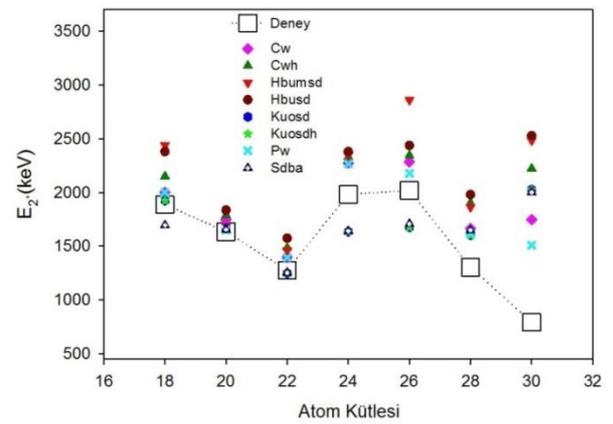

**Şekil 3.** Çift-çift Ne izotopları için diğer matris elemanları kullanılarak SM hesaplamalarından elde edilen ve deneysel $2^+$ enerji seviyeleri

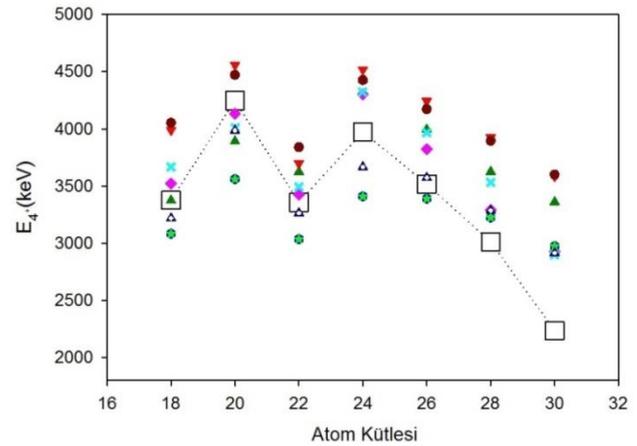

**Şekil 4.** Çift-çift Ne izotopları için diğer matris elemanları kullanılarak SM hesaplamalarından elde edilen ve deneysel $4^+$ enerji seviyeleri

$2^+$ ve $4^+$ enerji seviye değerlerinin birbirlerine oranları ($R_{4/2}$), çekirdeğin deformasyonu hakkında bilgi veren bir değerdir. Buna göre, $R_{4/2}$ oranının 2,00'dan küçük olması çekirdeğin kolektif yapıda olmadığına, 2,00 civarında olması küresel titreşici olduğuna, 2,50 civarında olması geçişken olduğuna ve 3,33 civarında olması da katı-rotor olduğuna işaret eder. Şekil 5'den görüldüğü gibi, [18]Ne ve ve [26]Ne'nin kolektif yapıda olmadığı görülmektedir. [20]Ne ve [22]Ne geçişken ve [24]Ne küresel titreşici olup, deneysel değerler ile teori birbiriyle uyum içerisindedir. [28]Ne ve 30Ne için ise, elde edilen teorik sonuçların, deneysel değerlerden uzak olduğu görülmektedir. Deneysel değerlere göre bu izotoplar geçişken karakterli izotoplardır. Oysa *usd* tipli matris eleman setleri ile yapılan teorik hesaplamalar sonucunda bunların küresel titreşici oldukları ya da kolektif yapıda olmadıkları görülmüştür. Diğer matris eleman setlerinden *kuosd*, 2,26 değeri ile, [28]Ne izotopu için deneysel olan 2,31 değerine en yakın sonucu vermiştir. [30]Ne izotopu için ise, 2,82 olan deneysel değere en yakın sonucu, 1,92 ile *pw* seti vermiştir.

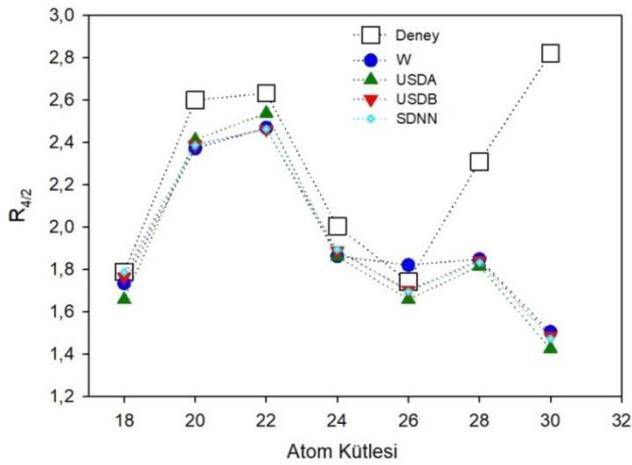

**Şekil 5.** Çift-çift Ne izotopları için ilk $2^+$ ve $4^+$ enerji seviye değerlerinin birbirine oranları

Taban durumdan ilk uyarılmış $2^+$ durumuna geçişin olasılığı, indirgenmiş geçiş olasılığı parametresi (*B(E2)*) hesaplanarak verilmiştir. Literatürde teorik ve deneysel sonuçlardan elde edilen pek çok değerin kullanılmasıyla elde edilen kabul edilmiş değerler alınarak, kabuk modeli hesaplamalarıyla karşılaştırılmıştır. Bu değer, çekirdeklerin kolektif davranışlarını görmede önemli bir niceliktir ve nükleer yapı hakkında bilgi verir. Şekil 6'da görüldüğü gibi, [20]Ne çekirdeği için *B(E2)* değerinin yüksek olması, yüksek kolektifliğe işaret etmektedir [20, 29].

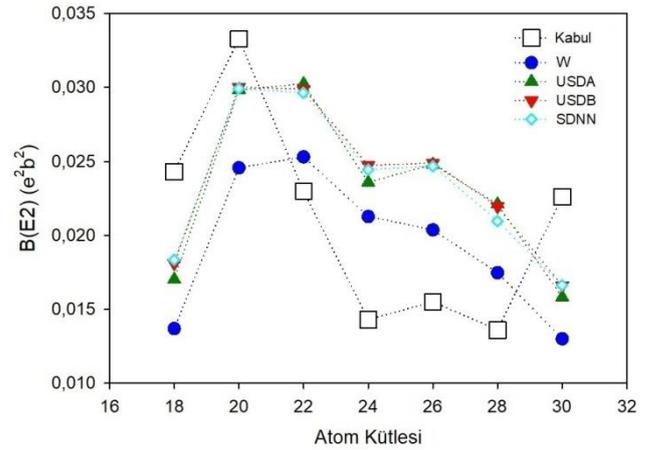

**Şekil 6.** Çift-çift Ne izotopları için *B(E2)* değerleri

*B(E2)* değerlerinden elde edilen kuadrupol deformasyon parametresi ($\beta_2$) ise, çekirdek deformasyonunun bir göstergesidir. Bu değerin pozitif olması, çekirdeğin şeklinin kutuplardan çekilmiş gibi olduğuna, negatif olması ise kutuplardan basılmış gibi olduğuna işaret eder. $\beta_2$ değeri ne kadar büyük ise, bu çekilme ya da basılma da o kadar fazladır. Bu fazlalık ise, çekirdeğin deformasyonunun derecesini gösterir. Şekil 7'den görülebileceği gibi tüm çift-çift Ne çekirdekleri, kutuplardan çekilmiş şekildedir. *Usda, usdb* ve *sdnn* matris eleman setleri ile yapılan hesaplamaların birbirlerine yakın oldukları ve [18]Ne, [20]Ne ve [30]Ne çekirdekleri için kabul edilmiş değerlere *w* etkileşme setinden daha yakın olduğu görülmektedir. Diğer Ne çekirdekleri için ise, *w* matris eleman seti ile yapılan hesaplamalardan elde edilen sonuçların kabul edilmiş değerlere [20, 29] daha yakın olduğu görülmektedir. Kabul edilmiş sonuçlara göre, bu izotoplar içinde deformasyonu en fazla olan, [20]Ne çekirdeğidir. Deformasyonu en az olanlar ise, [24]Ne, [26]Ne ve [28]Ne çekirdekleridir. Bu çalışmada yapılan teorik hesaplamalara göre ise, [20]Ne çekirdeğinin deformasyonu en fazla elde edilmiş olup literatürle uyumludur. Deformasyonu en az olan ise, [30]Ne çekirdeğidir.

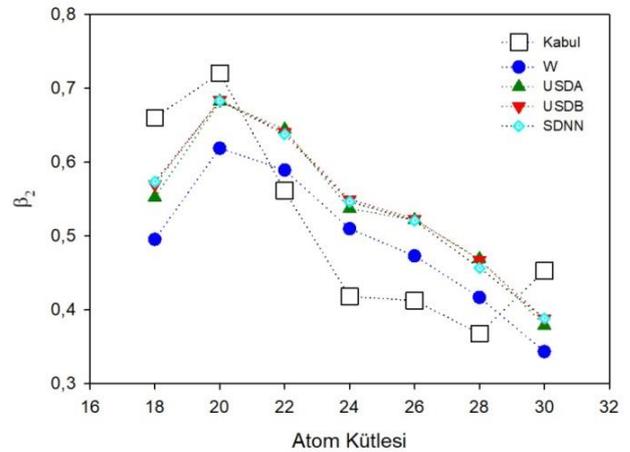

**Şekil 7.** Çift-çift Ne izotopları için $\beta_2$ değerleri

Son olarak, bu çalışmada ilk $6^+$ seviye enerjileri de bazı Ne izotopları için hesaplanabilmiştir. $^{22}$Ne çekirdeği için literatürde mevcut olan değerde kesinlik yoktur ve sadece *usdb* ve *sdnn* matris eleman setleri kullanılarak elde edilebilmiştir. $^{28}$Ne için literatürde değer bulunmamasına karşın, *hbumsd* matris eleman seti ile bir değer elde edilebilmiştir. Tablo 1'de, hesaplanabilen ilk $6^+$ seviye enerjileri, farklı matris eleman setleri için deneysel değerlerle birlikte verilmiştir.

**Tablo 1.** Deney ve faklı matris elemanları ile yapılan hesaplardan elde edilen ilk $6^+$ enerji seviye değerleri

| | First 6+ Energy (keV) | | | | | | |
|---|---|---|---|---|---|---|---|
| *İzotop* | *Deney* | *cw* | *cwh* | *hbumsd* | *hbusd* | *kuosd* | *kuosdh* |
| $^{20}$Ne | 8778 | 8546 | 7668 | 8560 | 9241 | 8188 | 7803 |
| $^{22}$Ne | 6311 | - | - | - | - | - | - |
| $^{28}$Ne | - | - | - | 6806 | - | - | - |
| | *pw* | *sdba* | *w* | *usda* | *usdb* | *sdnn* | |
| $^{20}$Ne | 8405 | 8564 | 8515 | 8360 | 8547 | 8593 | |
| $^{22}$Ne | - | - | - | - | 6244 | 6163 | |
| $^{28}$Ne | - | - | - | - | - | - | |

## 4. SONUÇ

Bu çalışmada, *sd* model uzayında $^{16}$O öz çekirdeğinin üzerinde bulunan çift-çift Ne çekirdeklerinin nükleer yapılarının araştırılması amaçlanmıştır. Bu amaçla, çekirdeklerin ilk uyarılmış seviye enerjileri, bu enerjilerin birbirlerine oranları, seviyeler arasındaki geçiş olasılıkları ve deformasyon parametreleri incelenmiştir. Nötron sayısının sihirli olduğu $^{18}$Ne çekirdeği için, ilk uyarılmış seviye enerjisinin beklendiği gibi fazla olduğu görülmüştür. $^{20}$Ne ve $^{22}$Ne çekirdekleri için bu uyarılma enerjisi düşmekte ve sonra $^{24}$Ne ve $^{26}$Ne çekirdekleri için tekrar yükselmektedir. Bunun sebebi bu çekirdekler için sırasıyla $d_{5/2}$ ve $s_{1/2}$ yörüngelerinin tam dolu hale gelmiş olmasıdır. $^{30}$Ne çekirdeği için ise $d_{3/2}$ yörüngesi tam dolu hale gelerek, kabuğun kapanmasına neden olacaktır. Bu nedenle, bu çekirdek için de ilk uyarılmış durumun enerjisinin yükselmesi beklenir. Bu çalışmada yaptığımız teorik hesaplamalarda bu yükselmeyi görmemize rağmen, literatürde mevcut ama kesin olarak belirlenmemiş olan deneysel verilere göre, bu artış gözlenmemektedir. Ancak, bu izotopun ilk uyarılmış enerjisinin bu derece düşük olması, küresel şekilde olması yerine oldukça deforme bir çekirdek olduğuna işaret eder [30]. Bu çekirdek, standart olmayan özellikler gösteren ve nötron damlama çizgisine oldukça yakın yerleşmiş inversiyon adası olarak adlandırılan bölgede yer alan bir çekirdektir [31]. Bu nedenle, bu çalışmada yapılan hesapların, standart olmayan bu bölge çekirdekleri açıklayamaması normaldir. Çekirdek deformasyonları incelendiğinde ise, $^{18}$Ne için deformasyonun olduğu olması, beklendiği gibi hesaplanmıştır. Yine beklendiği üzere, ardından gelen $^{20}$Ne ve $^{22}$Ne çekirdeklerinde deformasyonun, $^{18}$Ne çekirdeğine göre yüksek olduğu gözlenmiştir. $^{24}$Ne ve $^{26}$Ne için yine yörüngelerin dolması nedeniyle deformasyonun büyüklüğü azalmıştır. Deformasyon için literatürdeki kabul edilmiş verilerin davranışları incelendiğinde ise, sihirli sayıda nötron sahip olmasına rağmen, $^{30}$Ne çekirdeği için deformasyonun büyüdüğü görülmektedir.

## KAYNAKÇA